\documentclass{Interspeech}

\interspeechcameraready

\title{HuBERT-VIC: Improving Noise-Robust Automatic Speech Recognition of Speech Foundation Model via Variance-Invariance-Covariance Regularization}

\author{Hyebin}{Ahn}
\author{Kangwook}{Jang}
\author{Hoirin}{Kim}

\affiliation[nocounter]{School of Electrical Engineering}{KAIST}{Republic of Korea}
\email{ahb9900@kaist.ac.kr, dnrrkdwkd12@kaist.ac.kr, hoirkim@kaist.ac.kr}

\usepackage[table,xcdraw,dvipsnames]{xcolor}

\usepackage{bbm}
\usepackage{graphicx}
\usepackage{amsmath}
\usepackage{amssymb}
\usepackage{booktabs}
\usepackage{enumitem}
\usepackage{rotating}
\usepackage{colortbl}
\usepackage{makecell}
\usepackage{tabu}
\usepackage{subcaption}
\usepackage{multicol}
\usepackage{multirow}
\usepackage{csquotes}
\usepackage{wrapfig}
\usepackage{url}

\begin{document}

\maketitle

\begin{abstract}
Noise robustness in speech foundation models (SFMs) has been a critical challenge, as most models are primarily trained on clean data and experience performance degradation when the models are exposed to noisy speech. To address this issue, we propose HuBERT-VIC, a noise-robust SFM with variance, invariance, and covariance regularization (VICReg) objectives.
These objectives adjust the statistics of noisy speech representations, enabling the model to capture diverse acoustic characteristics and improving the generalization ability across different types of noise. When applied to HuBERT, our model shows relative performance improvements of 23.3\% on LibriSpeech \textit{test-clean} and 13.2\% on \textit{test-other}, compared to the baseline model pre-trained on noisy speech.

\end{abstract}

\section{Introduction}
Over the past few years, speech foundation models (SFMs) have significantly advanced the field of speech processing, achieving impressive improvements in automatic speech recognition (ASR)\,\cite{schneider2019wav2vec,baevski2020wav2vec,baevski2022data2vec,hsu2021hubert, ling2020decoar,chung2021w2v} and other speech-related tasks\,\cite{chen2021wavlm,jang2024star,liu2020mockingjay,liu2021tera,ng2022i2cr}.
SFMs such as wav2vec 2.0\,\cite{baevski2020wav2vec} and HuBERT\,\cite{hsu2021hubert} have demonstrated exceptional performance by leveraging large amounts of unlabeled speech data, reducing the reliance on annotated data.
A key factor behind the success of SFMs is a pretext scheme that predicts codewords of masked frames\,\cite{devlin-etal-2019-bert}, and this enables SFMs to capture the complex acoustic patterns from speech.
However, since these models are mainly trained on clean speech, their recognition capability is weakened when the models are exposed to noisy environments and this leads to severe performance degradation in real-world noisy scenarios.

Several studies have attempted to enhance the noise robustness of SFMs, employing contrastive learning\,\cite{zhu2022noise,wang2022wav2vec,zhu2023robust,zhu2022joint}, reconstruction loss\,\cite{wang2022improving,sadhu2021wav2vec} or knowledge distillation\,\cite{wang2023hubert} during the pre-training stage.
Wav2vec-Switch\,\cite{wang2022wav2vec} incorporates additional contrastive losses with clean-noisy switched contextualized targets into wav2vec 2.0 to enforce prediction consistency between the clean and noisy speech.
Another work\,\cite{wang2022improving} introduces a reconstruction module, where the model is trained to minimize the difference from clean speech using contrastive learning.
HuBERT-AGG\,\cite{wang2023hubert} learns noise-invariant representations by distilling clean teacher’s knowledge with layer-wise aggregation, making the model well-suited for ASR.

However, previous methods are generally vulnerable to representation collapse, as they rely heavily on joint embedding networks without explicit regularization to control the variability of feature channel dimensions\,\cite{bardes2022vicreg}. 
Another issue is that since previous methods are mostly based on contrastive loss\,\cite{wang2022wav2vec}, which requires large batch sizes and high resources, training become more difficult due to the slow convergence and various hyperparameter settings. There is an another work based on knowledge distillation\,\cite{wang2023hubert}. However, they require an additional ASR fine-tuning stage to measure each transformer layer's contribution to ASR for the effective distillation of the clean teacher's knowledge to the student.

From this perspective, we propose a Variance-Invariance-Covariance Regularization\,(VICReg)\,\cite{bardes2022vicreg}-based loss to effectively control the feature-level statistics on noise-robust pre-training of SFM.
VICReg consists of three distinct terms—Variance, Invariance, and Covariance—applied within the clean-noise knowledge distillation framework.
While the invariance term ensures that the noisy representations align with the clean representations as a reference, the variance term directly controls the statistics of noisy speech representations by regulating the variability of channel dimensions.
Additionally, the covariance term improves the quality of noisy speech representations by reducing redundancy between every pair of channel dimensions. This decorrelation enables the model to better capture diverse speech characteristics.
Simply adding VICReg terms improves performance in noisy environments, highlighting its potential as a noise-robust pre-training method.

Our contributions can be summarized as follows:\,(i) we propose \textbf{HuBERT-VIC}, a novel noise-robust pre-training method that integrates variance, invariance, and covariance regularization terms to pretext task of SFMs.
(ii) Experimental results conducted on LibriSpeech\,\cite{panayotov2015librispeech} with MUSAN\,\cite{snyder2015musan} noise added show that our proposed HuBERT-VIC surpasses other methods across various noise types and Signal-to-Noise Ratio (SNR) levels.
(iii) We analyze the relationship between variance, SNRs, and ASR performance, demonstrating how the variance of channel dimensions contributes to the noise robustness in SFMs.

\section{Method}

\begin{figure*}[hbt!]
  \centering
  \includegraphics[width=\linewidth]{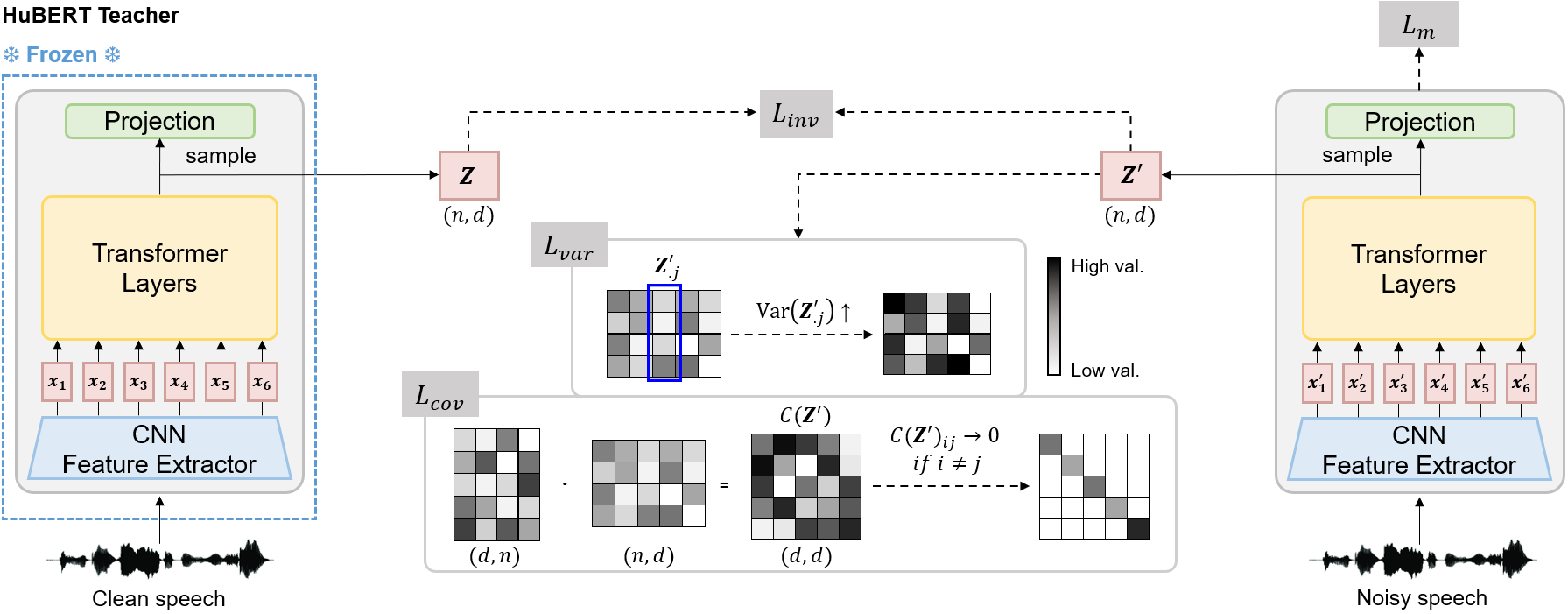}
  \caption{Overview of the proposed HuBERT-VIC training pipeline. Teacher model is pre-trained on clean speech and remains frozen during noise-robust pre-training of student model. The student model is initialized with the same clean pre-trained weights as the teacher, and then trained with noise-augmented speech inputs.
  Randomly selected frames of the final Transformer layer representations from both models are used to compute the VICReg loss, guiding the noise-robust pre-training.
  }
  \label{fig:my_framework}
  \vspace{-10pt}
\end{figure*}

\subsection{HuBERT}

HuBERT\,\cite{hsu2021hubert} is a SFM that employs the masked prediction task\,\cite{devlin-etal-2019-bert,taylor1953cloze}. In this task, segments of the input sequence are masked, and the model predicts the corresponding hidden unit codeword.
HuBERT consists of convolutional neural network (CNN) feature encoder followed by dozens of consecutive Transformer\,\cite{vaswani2017attention} layers. 
Given a length $T$ sequence of masked representations $\mathbf{X} = [\mathbf{x}_1, \mathbf{x}_2, ..., \mathbf{x}_T]$ from the CNN feature extractor with masked timesteps $M$, the masked prediction objective of HuBERT can be defined as,
\begin{equation}
    \mathcal{L}_m(\mathbf{X}) = \sum_{t \in M} \log p\left(\mathbf{c}_t | \mathbf{X}, t\right),
    \label{eq:hubertlm}
\end{equation}
where $\mathbf{C} = [\mathbf{c}_1, \mathbf{c}_2, ..., \mathbf{c}_T]$ is the probability distributions of $T$ timesteps over $C$ codeword candidates.

\subsection{HuBERT-VIC} 



To enhance the noise robustness of speech representations, we adapt the variance, invariance, and covariance regularization terms originally designed for joint embedding networks in computer vision domain\,\cite{bardes2022vicreg}. 
Incorporating all these regularization terms maintains sufficient variance of noisy speech representations, enforces invariance to noise, and prevents overfitting to specific noise characteristics. This approach improves the model's noise robustness and generalization ability.


%
Fig.\,\ref{fig:my_framework} shows the training pipeline of our proposed HuBERT-VIC. The teacher model is pre-trained on clean speech and remains frozen during noise-robust pre-training with clean speech inputs. The student model, initialized with the same clean pre-trained weights with the teacher's, is trained on noise-augmented speech inputs that retain the same linguistic information with the teacher's inputs.
We sample $n$ time frames along the time axis from the output of the final Transformer layer of both the teacher and student models. These frames are randomly selected from multiple utterances within a batch. Then $\mathbf{Z}, \mathbf{Z}' \in \mathbb{R}^{n \times d}$ represent the sampled representations for the teacher and student models, where $d$ represents the channel dimensions. These representations can be written as $\mathbf{Z} = [\mathbf{z_1}, \mathbf{z_2}, \dots, \mathbf{z_n}]$ and $\mathbf{Z}' = [\mathbf{z'_1}, \mathbf{z'_2}, \dots, \mathbf{z'_n}]$.
This sampling process makes the features more independent of consecutive time frames and promotes training stability, similar to a prior work\,\cite{ng2023hubert}.


\noindent \textbf{Invariance term}\quad Given the sampled representations $\mathbf{Z}$ and $\mathbf{Z}'$, the invariance regularization term minimizes the discrepancy between the teacher’s clean speech representations and the student’s noisy counterparts. %
Encouraging the invariance for the teacher and student is crucial for noise robustness, as it ensures that the model learns to maintain consistency in the representations even in the presence of noise. Then the model becomes more robust to varying noise conditions and better generalize across different noise types.
The invariance term is defined as the mean squared error (MSE),
\begin{equation}
    s(\mathbf{Z}, \mathbf{Z}') = \frac{1}{n} \sum_{i=1}^n \big|\big|\mathbf{z}_i - \mathbf{z}'_i\big|\big|^2_2.
    \label{eq:inv_equ}
\end{equation}

\vspace{3pt}
\noindent \textbf{Variance term}\quad The variance regularization ensures sufficient dispersion across the channel dimensions of $\mathbf{Z}'$ and prevents representation collapse where the learned representations become overly concentrated in certain dimensions. Based on the $j^{th}$ column vector $\mathbf{Z}'_{\cdot j}=[Z'_{1j}, Z'_{2j}, \dots, Z'_{nj}]^T$, the variance term is defined as follows,
\begin{equation}
    v(\mathbf{Z}') = \frac{1}{d} \sum_{j=1}^d \max(0, \gamma - \sqrt{\text{Var}(\mathbf{Z}'_{\cdot j}) + \epsilon}).
    \label{eq:var_equ}
\end{equation}
Each channel dimension maintains a minimum level of variability by applying a hinge loss with a hyperparameter $\gamma$, which controls the threshold, and a small scalar $\epsilon$ to prevent numerical instabilities.
The variance term ensures a balanced distribution of information across channel dimensions and induces the model to capture diverse essential acoustic characteristics for noise robustness.
%
Certain level of high variance in channel dimensions correlates with better handling of noisy speech, as it allows the model to capture a broader range of variations on speech (See Sec.\,\ref{sec:ablation}).
The variance term complements the invariance term by enabling the model to learn differences between clean and noisy representations. By capturing varied acoustic characteristics, this term can help the model to learn noise-specific features and reduce the impact of noise.

\vspace{3pt}
\noindent \textbf{Covariance term}\quad The covariance regularization aims to reduce redundancy between every pair of channel dimensions so that each channel dimension can capture distinct and independent information from the others.
Covariance matrix of noise-perturbed representations $C(\mathbf{Z}') \in \mathbb{R}^{d \times d}$ is calculated to capture relationships between channel dimensions.
The off-diagonal elements of $C(\mathbf{Z}')$ are then penalized with $c(\mathbf{Z}')$ by pushing toward zero.
\begin{equation}
    C(\mathbf{Z}') = \frac{1}{n-1} \sum_{i=1}^n (\mathbf{z}'_i - \mathbf{\bar{z}}')^T(\mathbf{z}'_i - \mathbf{\bar{z}}'),  \ \mathbf{\bar{z}}'=\frac{1}{n} \sum_{i=1}^n \mathbf{z}_i'.
\label{eq:cov_equ}
\end{equation}
\begin{equation}
    c(\mathbf{Z}') = \frac{1}{d} \sum_{i\neq j} \big|\big|C(\mathbf{Z}')_{ij}\big|\big|_2^2.
    \label{eq:cov_offdiag}
\end{equation}
The covariance term enhances the noise robustness by reducing redundancy across channel dimensions, enabling the model to capture more distinct and diverse features of speech and improving its ability to generalize well to various environments.

By combining these three regularization terms, the VIC loss is defined as follows,
\begin{equation}
    \mathcal{L}_{VIC} = \lambda s(\mathbf{Z},\mathbf{Z}') + \mu v(\mathbf{Z}') + \nu c(\mathbf{Z}').
\label{eq:vic_equ}
\end{equation}
\begin{equation}
    \mathcal{L}_{tot} = \mathcal{L}_m + \alpha \mathcal{L}_{VIC}.
\label{eq:total_equ}
\end{equation}
The combined final objective $\mathcal{L}_{tot}$ consists of regularization loss $\mathcal{L}_{VIC}$ and the masked prediction loss $\mathcal{L}_m$ of HuBERT.
These components are jointly optimized during the noise-robust pre-training of HuBERT.

\section{Experiments}

\subsection{Dataset and Evaluation Descriptions}
For noise-robust pre-training, we utilize unlabeled 960 hours of LibriSpeech\,\cite{panayotov2015librispeech} and MUSAN\,\cite{snyder2015musan} (\textit{babble}, \textit{music}, and \textit{natural}) with SNRs uniformly sampled between 5 to 10dB.
Pre-processing for MUSAN dataset, such as dataset partitioning and \textit{babble} noise synthesis, follows the pipeline used in\,\cite{shi2022robust}.
Then we evaluate on LibriSpeech exploiting all noise categories of MUSAN dataset with discrete SNR levels of $\{0,5,10,15\}$dB to assess the model performance across a wide range of noise conditions (Tab.\,\ref{tab:main_result_lrs2_snr015}).

For direct comparison with prior work\,\cite{wang2023hubert}, we conduct another experiment following a protocol where LibriSpeech is augmented using only MUSAN \textit{music} and \textit{natural} noise with random SNR levels between 5 and 10dB for pre-training (Tab.\,\ref{tab:mainres_wobabble}).
In this case, the SNR level for noise augmentation during evaluation is also randomly chosen in the same range.
In both settings, we fine-tune our model on labeled \textit{train-clean-100} subset of LibriSpeech without noise perturbation to improve core ASR capabilities\,\cite{wang2022wav2vec,wang2023hubert}.
Word error rate (WER) and average WER across different noise types, N-WER\,\cite{shi2022robust, kim2024learning} are reported for ASR performance.



\subsection{Model and Training Descriptions}
\noindent \textbf{Noise-robust pre-training}\quad Both the teacher and student models are initialized with officially released HuBERT-\textsc{Base} checkpoint, which consists of 7 layers of 1D convolution and 12 Transformer\,\cite{vaswani2017attention} layers with embedding dimension $d$ as 768.
Masking and dropouts are turned off in the teacher model to distill clear representations to the student model.
We implement the masked prediction objective of HuBERT with default configuration in Fairseq recipe\,\cite{ott2019fairseq} for 50k training steps, applying Adam optimizer and pre-trained $k$-means clustering model with $k=500$.
For regularization terms, we set $\lambda$ as 5, $\mu$ and $\nu$ as 1, respectively, for each V-I-C weight in Eq.\,\ref{eq:vic_equ}.
We adopt $\gamma$ as 1, $\epsilon$ as 1e-4 in Eq.\,\ref{eq:var_equ}, and $\alpha$ for 1 in Eq.\,\ref{eq:total_equ}.
The number of randomly selected samples $n$ is set to 512, and the same samples are selected for both the teacher and student models.

\vspace{3pt}
\noindent \textbf{ASR fine-tuning and decoding}\quad We follow \textit{base\_100h} configuration from Fairseq, which adds a linear layer after the last Transformer layer and optimizes with Connectionist Temporal Classification (CTC) loss\,\cite{graves2006connectionist}.
The teacher model is discarded and only noise-robust pre-trained student model is fine-tuned.
For decoding, we adopt a publicly available 4-gram language model trained on LibriSpeech with a beam size of 1500.
We set the other decoding-related hyperparameters based on \textit{infer\_kenlm} configuration from Fairseq.

\section{Results}

\subsection{Main Results}
\begin{table*}[!t]
    \centering
    \footnotesize
    \caption{WER (\%) of noise-robust pre-trained models on LibriSpeech \textit{test-clean} and \textit{test-other}, with MUSAN \textit{Babble}, \textit{Music}, and \textit{Natural} subsets augmented at SNRs ranging from 0 to 15dB. PT Type refers to whether the model is trained on clean speech or noise-augment speech. HuBERT-AGG${}^{\dag}$ is reproduced.}
    \label{tab:main_result_lrs2_snr015}
    \vspace{-5pt}
    \addtolength{\tabcolsep}{-0.5pt}
    \renewcommand{\arraystretch}{1.1}
    \resizebox{\textwidth}{!}{
    \begin{tabular}{lc|ccccc|ccccc|ccccc|c|c}
    \toprule
    \multirow{2}{*}{Method} & \multirow{2}{*}{PT Type} & \multicolumn{5}{c|}{Babble, SNR (dB) $=$} & \multicolumn{5}{c|}{Music, SNR (dB) $=$} & \multicolumn{5}{c|}{Natural, SNR (dB) $=$} & \multicolumn{1}{c|}{N-WER} & Clean \\
    & & 0 & 5 & 10& 15 & AVG & 0 & 5 & 10 & 15 & AVG & 0 & 5 & 10 & 15  & AVG & AVG\! & $\infty$ \\
    \midrule
    \multicolumn{5}{l}{\textit{\textcolor{gray}{test-clean eval.}}}  \\    
    HuBERT\,\cite{hsu2021hubert} & clean & 60.7 & 20.6 & 7.3 & 4.6  & 23.3 & 35.0 & 15.1 & 6.6 & 4.5 & 15.3 & 22.8 & 10.1 & 5.4 & 4.1 & 10.6 & 16.4 & 3.4 \\
    HuBERT\,\cite{hsu2021hubert} & noisy & 17.2 & 7.4 & 5.3 & 4.7 &  8.7 & 11.2 & 6.3 & 5.0 & 4.6 & 6.8 & 9.8 & 6.1 & 5.1 & 4.7 & 6.4 & 7.3 & 4.3 \\
    HuBERT-AGG${}^{\dag}\,\cite{wang2023hubert}$ & noisy & 14.6 & 6.2 & 4.6 & 4.2 & 7.4 & 9.4 & 5.3 & 4.4 & 4.1 & 5.8 & 8.3 & 5.3 & 4.5 & 4.1 & 5.6 & 6.3 & 3.7 \\
    \cellcolor[HTML]{f9caf1}\textbf{Ours} & \cellcolor[HTML]{f9caf1}noisy & \cellcolor[HTML]{f9caf1}\textbf{13.4} & \cellcolor[HTML]{f9caf1}\textbf{5.7} & \cellcolor[HTML]{f9caf1}\textbf{4.1} & \cellcolor[HTML]{f9caf1}\textbf{3.7} & \cellcolor[HTML]{f9caf1}\textbf{6.7} & \cellcolor[HTML]{f9caf1}\textbf{8.6} & \cellcolor[HTML]{f9caf1}\textbf{4.7} & \cellcolor[HTML]{f9caf1}\textbf{3.9} & \cellcolor[HTML]{f9caf1}\textbf{3.7} & \cellcolor[HTML]{f9caf1}\textbf{5.2} & \cellcolor[HTML]{f9caf1}\textbf{7.5} & \cellcolor[HTML]{f9caf1}\textbf{4.7} & \cellcolor[HTML]{f9caf1}\textbf{3.9} & \cellcolor[HTML]{f9caf1}\textbf{3.6} & \cellcolor[HTML]{f9caf1}\textbf{4.9} & \cellcolor[HTML]{f9caf1}\textbf{5.6} & \cellcolor[HTML]{f9caf1}\textbf{3.4} \\
    \midrule
    \multicolumn{5}{l}{\textit{\textcolor{gray}{test-other eval.}}}  \\
    HuBERT\,\cite{hsu2021hubert} & clean & 79.4 & 44.4 & 21.8 & 12.9 & 39.6 & 53.3 & 32.1 & 18.1 & 12.1 & 28.9 & 39.6 & 24.1 & 15.2 & 11.2 & 22.5 & 30.4 & 8.2 \\
    HuBERT\,\cite{hsu2021hubert} & noisy & 38.7 & 19.8 & 13.3 & 10.2 & 20.5 & 26.3 & 16.1 & 12.1 & 10.8 & 16.3 & 23.2 & 15.5 & 12.1 & 10.7 & 15.4 & 17.4 & 9.5 \\
    HuBERT-AGG${}^{\dag}$\,\cite{wang2023hubert} & noisy & 35.0 & 17.1 & 11.7 & 9.9 & 18.4 & 23.0 & 13.9 & 10.7 & 9.4 & 14.3 & 20.1 & 13.3 & 10.6 & 9.4 & 13.4 & 15.4 & 8.6 \\
    \cellcolor[HTML]{f9caf1}\textbf{Ours} & \cellcolor[HTML]{f9caf1}noisy & \cellcolor[HTML]{f9caf1}\textbf{34.8} & \cellcolor[HTML]{f9caf1}\textbf{16.9} & \cellcolor[HTML]{f9caf1}\textbf{11.4} & \cellcolor[HTML]{f9caf1}\textbf{9.5} & \cellcolor[HTML]{f9caf1}\textbf{18.2} & \cellcolor[HTML]{f9caf1}\textbf{22.8} & \cellcolor[HTML]{f9caf1}\textbf{13.5} & \cellcolor[HTML]{f9caf1}\textbf{10.4} & \cellcolor[HTML]{f9caf1}\textbf{9.1} & \cellcolor[HTML]{f9caf1}\textbf{14.0} & \cellcolor[HTML]{f9caf1}\textbf{20.0} & \cellcolor[HTML]{f9caf1}\textbf{13.2} & \cellcolor[HTML]{f9caf1}\textbf{10.5} & \cellcolor[HTML]{f9caf1}\textbf{9.3} & \cellcolor[HTML]{f9caf1}\textbf{13.3} & \cellcolor[HTML]{f9caf1}\textbf{15.1} & \cellcolor[HTML]{f9caf1}\textbf{8.1} \\
    \bottomrule
    \end{tabular}
    }
\vspace{-10pt}
\end{table*}

\renewcommand{\arraystretch}{0.95}
\begin{table}[hbt!]
    \centering
    \caption{WER (\%) on MUSAN Music, Natural subsets. Noisy speech is synthesized with random SNRs from 5 to 10dB. PT Type refers to whether the model is trained on clean speech or noise-augment speech. The results of the other methods are taken from previous work\,\cite{wang2023hubert}.}
    \vspace{-5pt}
    \label{tab:mainres_wobabble}
    \setlength{\tabcolsep}{5pt}
    \resizebox{\columnwidth}{!}{
        \begin{tabular}{lc|cc|cc}  
        \toprule
        \multirow{2}{*}{Method}         & \multirow{2}{*}{PT Type} & \multicolumn{2}{c|}{SNR\,(dB) $=\infty$} & \multicolumn{2}{c}{SNR\,(dB) $=5\text{--}10$}   \\
                                        & & \multicolumn{1}{c}{\footnotesize \textit{test-clean}} & \multicolumn{1}{c|}{\footnotesize \textit{test-other}} & \multicolumn{1}{c}{\footnotesize \textit{test-clean}} & \multicolumn{1}{c}{\footnotesize \textit{test-other}} \\ \midrule
        HuBERT\,\cite{hsu2021hubert}                          & clean                           & 3.4            & 8.2            & 8.7            & 21.2           \\
        HuBERT\,\cite{hsu2021hubert}                          & noisy                           & 4.1            & 9.0            & \multicolumn{1}{c}{5.4}            & 14.1           \\ 
        HuBERT-NIT\,\cite{wang2023hubert}                      & noisy                           & 3.7            & 8.7            & \multicolumn{1}{c}{4.6}            & 11.7           \\ 
        HuBERT-AGG\,\cite{wang2023hubert}                      & noisy                           & 3.5            & 8.5            & \multicolumn{1}{c}{4.3}            & 11.6           \\ \midrule
        \textbf{Ours}             & noisy                           & \textbf{3.4}   & \textbf{8.1}   & \textbf{4.2}   & \textbf{11.3}  \\ \bottomrule
        \end{tabular}
    }
\vspace{-5pt}
\end{table}

\renewcommand{\arraystretch}{0.95}
\begin{table}[hbt!]
    \centering
    \caption{WER (\%) on MUSAN Music, Natural subsets for ablation. Noisy speech is synthesized with random SNRs from 5 to 10dB. The results of the first row${}^{\dag}$ are taken from previous work\,\cite{wang2023hubert}.}
    \vspace{-5pt}
    \label{tab:ablation}
    \setlength{\tabcolsep}{5pt}
    \resizebox{0.95\columnwidth}{!}{
        \begin{tabular}{l|cc|cc}  
        \toprule
        \multirow{2}{*}{Loss}          & \multicolumn{2}{c|}{SNR\,(dB) $=\infty$} & \multicolumn{2}{c}{SNR\,(dB) $=5\text{--}10$}   \\
                                    & \multicolumn{1}{c}{\footnotesize \textit{test-clean}} & \multicolumn{1}{c|}{\footnotesize \textit{test-other}} & \multicolumn{1}{c}{\footnotesize \textit{test-clean}} & \multicolumn{1}{c}{\footnotesize \textit{test-other}} \\ \midrule
        $\mathcal{L}_m{}^{\dag}$                                                     & 4.1            & 9.0            & 5.4            & 14.1           \\
        ~~(+) inv.                                                     & 3.5            & 8.2            & 4.4            & 11.7           \\ 
        ~~~~(+) var.                                                 & 3.5            & 8.2            & 4.3            & 11.5           \\  \midrule
        ~~~~~~\textbf{(+) cov.} \textbf{(Ours)}                      & \textbf{3.4}            & \textbf{8.1}            & \textbf{4.2}            & \textbf{11.3}           
        \\ \bottomrule
        \end{tabular}
    }
\vspace{-12.5pt}
\end{table}

We compare the WER evaluated under different noise types and SNR levels in Tab.\,\ref{tab:main_result_lrs2_snr015} with the original HuBERT and previous work.
When comparing the overall performance across all noise types by N-WER, our model achieves the lowest N-WER of 5.6\% on \textit{test-clean}, outperforming all other models. This includes a relative N-WER improvement of 23.3\%, with relative AVG WER improvements across three noise types for 23.0\%-23.5\% compared to HuBERT with noisyPT. On \textit{test-other}, our model also achieves the lowest N-WER of 15.1\%, demonstrating a relative improvement of 13.2\%, with average WER reductions of 11.2\%-14.1\% across the three noise types.

While the clean teacher's knowledge is distilled to the student through invariance regularization, the variance regularization helps maintain robust feature representations by capturing the inherent characteristics in the noisy speech. The covariance regularization further enhances the model's ability to learn the relationships between features' channel dimensions. 
Together, these terms help the model to exhibit strong generalization ability across various noise types, consistently outperforming existing models on both \textit{test-clean} and \textit{test-other} on diverse noise conditions.
Furthermore, unlike previous methods which experienced performance degradation on clean speech compared to the original HuBERT, our model effectively prevents this degradation. We have confirmed the effectiveness of our method in terms of generalization ability on clean speech. 

We also compare the performance of our proposed method with an experimental protocol of prior work\,\cite{wang2023hubert}, as shown in Tab.\,\ref{tab:mainres_wobabble}.
While HuBERT pre-trained on noisy speech enhances noise robustness, it underperforms on clean speech compared to its clean pre-trained counterpart.
HuBERT-NIT and HuBERT-AGG show further improvements in noise robustness while they also cause slight performance degradations on clean speech.
In contrast, HuBERT-VIC achieves the best performance under noisy conditions without sacrificing clean speech performance.
This demonstrates the efficacy of our method in clean speech, which achieves performance improvements through simple regularization terms without the need for additional training stages or preliminary ASR fine-tuning.

\subsection{Ablation and Analysis}\label{sec:ablation}
\textbf{Regularization term ablation}\quad To analyze the contribution of three V-I-C terms in noise-robust ASR, we conduct an ablation study by selectively applying the regularization terms and evaluate the model’s performance on both clean\,(SNR$=$$\infty$) and noisy\,(SNR$=5\text{--}10$) speech.
We follow the same experimental setup as Tab.\,\ref{tab:mainres_wobabble} for faster convergence of the experiments.

The results of this ablation study are presented in Tab.\,\ref{tab:ablation}.
Compared to the model only trained with the masked prediction loss $\mathcal{L}_m$, adding the invariance term shows significant performance improvements on both clean and noisy speech in the second row.
The invariance term is effective as it reduces the discrepancy between clean and noisy speech representations more directly and encourages the model to learn clean phonetic features regardless of the noise.
By applying the invariance term after timestep sampling, we can better match the scale of the regularization terms and improve the model's convergence, leading to further performance improvements.
Adding the variance term results in slightly better performance as it allows the model to capture a broader spectrum of features on each channel and learn diverse features related to noise. By those characteristics, the variance term supports the invariance term by reducing the impact of noise and helps the model maintain consistency between clean and noisy representations. 
Finally, adding the covariance term improves the ASR performance by reducing redundancy across channel dimensions, which allows the model to capture more distinct relationships between features. The covariance term also supports the invariance term by improving the alignment between clean and noisy representations. 
When all three terms are applied, the model achieves the best WER of 3.4\%, 8.1\% on clean speech, 4.2\%, 11.3\% on noisy speech.
These results show that while the masked prediction loss and invariance term are fundamental for noise-robust pre-training, the variance and covariance terms complement these losses and contribute to the superior performance.
%

\begin{figure}
     \centering
     \begin{subfigure}[b]{0.23\textwidth}
         \centering
         \includegraphics[width=\textwidth]{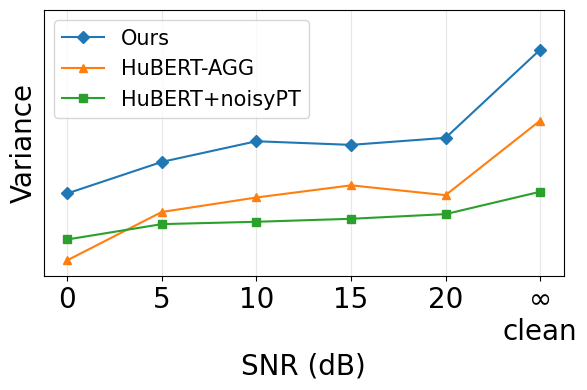}
         \caption{Result on babble noise.}
         \label{fig:var_babble}
     \end{subfigure}
     \begin{subfigure}[b]{0.23\textwidth}
         \centering
         \includegraphics[width=\textwidth]{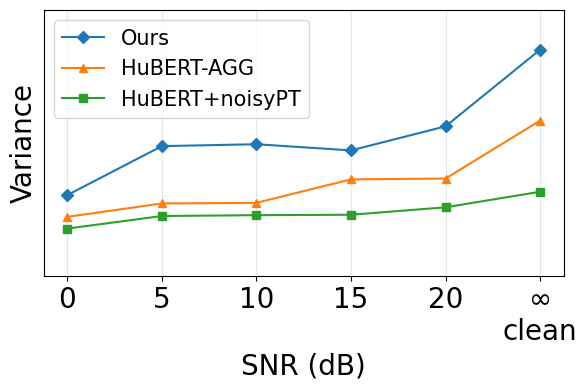}
         \caption{Result on music noise.}
         \label{fig:var_music}
     \end{subfigure}
     \vspace{-4pt}
        \caption{Variance of the channel dimensions in the last transformer output representations from the noise-robust pre-trained models, analyzed across different SNR levels.}
        \label{fig:var_analysis}
\vspace{-13pt}
\end{figure}

\vspace{3pt}
\noindent\textbf{Analysis on variance term}\quad We conduct an analysis to gain insights from the statistics of noisy speech representations in noise-robust pre-trained models, examining the relationship between channel dimensions' variance, SNR and ASR performance.
Fig.\,\ref{fig:var_analysis} shows various models' channel dimension variances of the last Transformer layer output representations.
The noisy speech inputs are based on LibriSpeech \textit{test-clean} with a mixture of MUSAN \textit{Babble}, \textit{Music} noise.
The x-axis represents the SNR levels, illustrating how the variance changes under different noise levels.
There is a trend in the variance of noisy speech representations; Higher SNR in input speech leads to higher variance in the channel dimensions of representations from noise-robust models.
%
We have observed that higher variance is present in higher SNR settings as shown in the Fig.\,\ref{fig:var_babble} and Fig.\,\ref{fig:var_music}. As SNR increases, the influence of noise diminishes and the model can capture diverse speech-related acoustic and phonetic features more than less-diverse noise features. Based on the fact that ASR performance is better at high SNR compared to low SNR, the increase in the variance can be interpreted as the model's enhanced ability to distinguish important speech characteristics, and ultimately its improved ASR performance at the end. These findings show that analyzing noisy speech statistics and applying the variance regularization is a simple yet effective approach for improving the noise robustness of SFMs.

\vspace{-1pt}
\section{Conclusion}
In this paper, we propose HuBERT-VIC for noise-robust ASR. The model consists of the invariance term, which learns knowledge from a teacher model to align clean and noisy speech representations, the variance term, which ensures sufficient variability across channel dimensions, and the covariance term, which reduces redundancy between different channel dimensions. These terms are incorporated into the HuBERT pre-training process to enhance the model's robustness to noisy speech. Experimental results demonstrate that our approach shows significant performance improvements compared to existing methods across various noise types and conditions, while also preventing performance degradation on clean speech and highlighting strong generalization ability.

\newpage

\section{Acknowledgements}
This work was conducted by Center for Applied Research in Artificial Intelligence(CARAI) grant funded by DAPA and ADD (UD190031RD).

\bibliographystyle{IEEEtran}
\bibliography{mybib}

\end{document}